%% file: main.tex
\documentclass{article} 
\usepackage{iclr2027_conference,times}

\usepackage[hyphens]{url}

\input{commands}

\usepackage[hidelinks]{hyperref}

\title{Interactive Clarification for Cloud \\ Infrastructure-as-Code Synthesis}

\author{Zhenning Yang, Kaden Gruizenga, Tongyuan Miao, Patrick Tser Jern Kon, Ang Chen \\
University of Michigan\\
\texttt{\{znyang,kgruiz,tymiao,patkon,chenang\}@umich.edu} \\
\And
Hui Guan, Andrew Barto \\
University of Massachusetts Amherst \\
\texttt{\{huiguan,barto\}@cs.umass.edu}
}

\iclrfinalcopy 

\begin{document}

\maketitle
\lhead{Preprint.}

\begin{abstract}

\input{abstract3}

\end{abstract}

\input{intro5}

\input{method8}
\input{evaluation6}
\input{related_work2}

\input{conclusion}
\bibliography{colm2026_conference,nsync_refs}
\bibliographystyle{iclr2027_conference}

\clearpage
\appendix
\input{appendix}


\end{document}

%% file: commands.tex
\usepackage{comment}
\usepackage{amsmath}
\usepackage{amsfonts}
\usepackage{amssymb}
\usepackage[T1]{fontenc}
\usepackage{textcomp}

\usepackage{graphicx}
\usepackage{booktabs}
\usepackage{multirow}
\usepackage{array}
\usepackage{makecell}

\usepackage{enumitem}
\usepackage{xcolor}
\usepackage{colortbl}   
\usepackage{pifont}
\usepackage{caption}

\usepackage{algorithm}
\usepackage{algorithmic}

\usepackage{listings}
\usepackage{tcolorbox}
\tcbuselibrary{skins,breakable,listings}

\setlist[itemize]{leftmargin=1.5em, topsep=4pt, itemsep=3pt, parsep=0pt, partopsep=0pt}
\setlist[enumerate]{leftmargin=1.5em, topsep=4pt, itemsep=3pt, parsep=0pt, partopsep=0pt}

\definecolor{downred}{rgb}{0.75,0.10,0.10}
\definecolor{upgreen}{rgb}{0.00,0.50,0.15}
\newcommand{\dnar}[1]{\textcolor{downred}{$\downarrow$#1}}
\newcommand{\upar}[1]{\textcolor{upgreen}{$\uparrow$#1}}

\newcommand{\graycell}{\cellcolor{gray!20}}

\newcommand{\stepnum}[1]{#1.}

\newcounter{obscount}

\newcommand{\mypar}[1]{{\textbf{#1.}\/}}

\newcommand{\stdd}[1]{{\small\,$\pm$#1}}

\newcommand{\ci}[2]{{\small\,[#1,\,#2]}}

\newcommand{\rtype}{\tau}
\newcommand{\Rset}{\mathcal{R}}
\newcommand{\Tset}{\mathcal{T}}
\newcommand{\Aset}{\mathcal{A}}
\newcommand{\code}[1]{\texttt{#1}}

\lstdefinestyle{promptstyle}{%
  basicstyle=\ttfamily\footnotesize,
  breaklines=true,
  breakatwhitespace=true,
  breakindent=0pt,
  columns=fullflexible,
  keepspaces=true,
  upquote=true,
  literate={—}{{\textemdash}}1,
}
\newtcblisting{promptbox}[1]{%
  enhanced,
  breakable,
  listing only,
  listing options={style=promptstyle},
  colback=black!4,
  colframe=black!50,
  colbacktitle=black!10,
  coltitle=black,
  fonttitle=\small\bfseries,
  boxrule=0.4pt, arc=1.5pt,
  left=3pt, right=3pt, top=2pt, bottom=2pt,
  before skip=6pt, after skip=6pt,
  title={#1},
}

%% file: abstract3.tex

The scale and complexity of modern cloud infrastructure have made ``Infrastructure-as-Code'' (IaC) essential for managing deployments through declarative configurations.
While large language models (LLMs) are increasingly used to generate IaC configurations from natural language, user requests are often ambiguous and underspecified. 
Unlike traditional code generation, it is costly and time-consuming to test IaC configurations during the synthesis, forcing the LLMs into an almost one-shot regime.
We observe that ambiguity in IaC synthesis exhibits a compositional structure: configurations decompose into three axes (resources, topology, attributes) where higher-level decisions constrain lower-level ones.
We propose a training-free, multi-level disambiguation framework that generates diverse candidate specifications, identifies structural disagreements across these axes, ranks them by informativeness, and produces targeted clarification questions that progressively narrow the configuration space.
We further introduce \textsc{Ambig-IaC}, an expert-verified benchmark of 300 validated IaC tasks with ambiguous requests, and define evaluation metrics based on graph edit distance and exact attribute matching. 
Comprehensive experiments show that our method outperforms existing interactive clarification baselines, with gains that scale with the interaction budget and generalize across models. 
Extensive ablation studies and analyses further demonstrate its robustness for interactive IaC synthesis.


%% file: intro5.tex
\section{Introduction}
\label{sec:intro}

\begin{figure}[t]
    \centering
    \includegraphics[width=0.75\columnwidth]{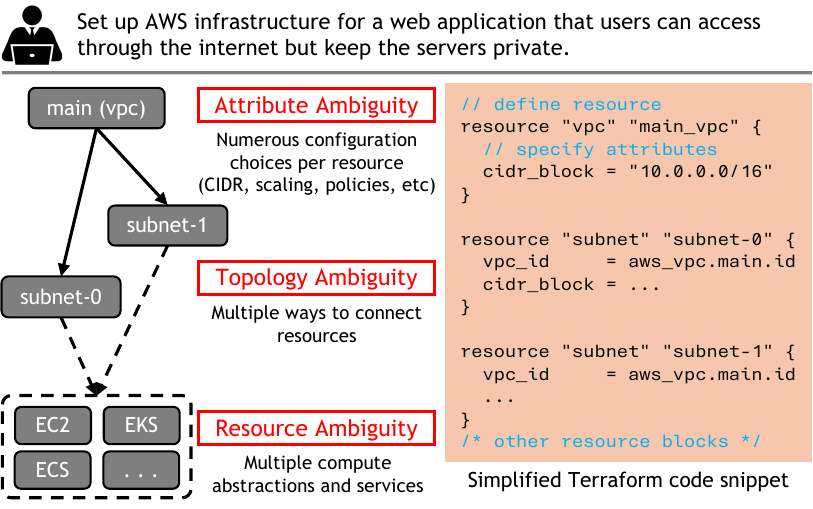}
    \caption{An underspecified user request can be satisfied by many plausible cloud infrastructure configurations, each represented as a resource dependency graph (left). Ambiguity arises in choosing resource abstractions and services for compute, networking, storage, in inter-resource topology, and in per-resource configuration attributes,  when translating intent into Infrastructure-as-Code (right).}
    \label{fig:intro-iac-code}
\end{figure}

LLMs are increasingly adopted for synthesizing structured artifacts from natural-language descriptions: code, database queries, and configuration programs~\citep{ambisql, tomswe, clarifygpt}.
However, user descriptions that drive this synthesis are often underspecified---prompts state high-level goals, not implementation details, so many valid outputs can satisfy the same prompt.
A model that silently ``resolves'' this ambiguity via its own priors may produce plausible but misaligned results.
Cloud Infrastructure-as-Code (IaC) is a particularly demanding instance of this problem~\citep{cloudagent_vision}.
IaC frameworks such as Terraform~\citep{terraform} define cloud infrastructure as declarative configurations~\citep{cloudagent_vision}, and recent work has begun generating them directly from natural language~\citep{iaceval}.
A single user request typically specifies only high-level infrastructure goals, leaving lower-level decisions (e.g., compute abstractions, resource connectivity, and security or networking policies) underspecified. This ambiguity gives rise to a large space of plausible yet potentially misaligned configurations (Figure~\ref{fig:intro-iac-code}).

One way to handle ambiguity is by iterative clarification and refinement. 
In conventional code generation, agents refine programs using execution feedback: running tests, observing failures, and repairing.
Cloud configuration offers no such loop.
Executing an IaC program requires provisioning real infrastructure, which is slow, costly, and often irreversible or subject to strict rate limits.
Dry-run validators such as \texttt{terraform plan} catch syntactic and type errors but cannot detect semantic misalignment with user intent.
Indeed, we show that roughly 90\% of generated configurations pass Terraform validation regardless of intent alignment (Section~\ref{subsec:ablation}).
In one recent public incident, an AI-assisted Terraform workflow deleted a production database, requiring roughly 24 hours of recovery~\citep{grigorev2026droppeddb}.
Cloud agents, therefore, operate in an almost one-shot regime where incorrect decisions incur real cost and are difficult to undo.
Hence, precise disambiguation is a first-class requirement in IaC generation. 

\looseness=-1
Existing clarification methods do not transfer readily to cloud IaC because they target different ambiguity spaces.
First, prior work primarily studies open-domain questions that rely on common-sense reasoning and admit a small set of salient interpretations~\citep{ambigqa,stargate,l2c}. In such settings, clarification questions are relatively straightforward to formulate, and ambiguity can often be resolved within two to three turns~\citep{zhang2025modelingfutureconversationturns,stargate,clarifygpt}.
IaC instead exhibits compositional ambiguity across resources, topology, dependencies, and attributes, producing a vast space of plausible configurations.
Second, methods from code generation rely on \emph{execution feedback}. 
They identify ambiguity through behavioral divergence among candidate programs on synthesized tests~\citep{clarifygpt,ITDCG}.
This is difficult for IaC because execution may provision or modify real cloud resources, so ambiguity must be resolved before deployment.
Third, existing approaches provide little guidance on where to direct clarification effort. 
A recent work~\citet{Sanidhya-2025-interact} confirm the consequence empirically: without structural guidance, LLMs waste their question budget on low-value clarifications. 
As Figure~\ref{fig:intro-iac-code} illustrates, IaC synthesis explores a structured configuration space where resources, topology, and attributes are tightly coupled, which we leverage to guide the clarification process.

We observe that IaC ambiguity comes from three axes that arise from the hierarchical nature of cloud configuration (\emph{resources}: which components to provision; \emph{topology}: how they connect; \emph{attributes}: what settings to apply): resource decisions constrain which topologies are feasible, and topology decisions constrain which attributes are relevant.
Because each axis admits multiple valid choices, we can probe the extent of ambiguity by generating diverse candidate specifications and comparing them structurally: points where candidates disagree reveal where ambiguity concentrates.
We propose a multi-level, disagreement-driven disambiguation framework that exploits this observation to guide clarification under a given interaction budget.
The framework generates diverse candidate specifications, identifies structural disagreements across the three axes, ranks them by informativeness, and uses the most informative disagreements to generate targeted clarification questions. 
After the user answers each question,  incompatible candidates are pruned, and the process repeats, progressively narrowing the configuration space.
The framework requires no training or fine-tuning and works with any sufficiently capable LLM.

This paper makes the following contributions:
\begin{itemize}
    \item We present the first investigation of interactive cloud IaC synthesis that exploits structure of IaC configurations to guide multi-turn user intent clarification.

    \item We propose a training-free, multi-level framework that decomposes IaC ambiguity into three axes (resource, topology, and attribute), generates diverse candidate specifications, and uses structural disagreements to produce targeted clarification questions that efficiently narrow the configuration space.

    \item We introduce \textsc{Ambig-IaC}, an expert-verified benchmark of 300 IaC tasks, with validated reference configurations and ambiguous prompts. We additionally propose an evaluation framework that measures configuration correctness along structure (via graph edit distance) and concrete resource--attribute values (via exact, occurrence-aware matching). 

    \item Our evaluation shows that our method consistently outperforms baselines across interaction budgets and generalizes across language models, at a fraction of the computational cost of information-gain alternatives. Our ablation studies further demonstrate its robustness.
    
\end{itemize}

%% file: method8.tex
\section{Motivation}
\label{sec:the-motivation}

Modern compute increasingly relies on large-scale cloud deployments, and this footprint continues to grow: Gartner forecasts worldwide IT spending of \$6.15 trillion in 2026, driven in part by hyperscale cloud demand~\citep{gartner-it-2026}. As cloud estates expand, they become harder to manage: 65\% of practitioners report increased cloud complexity over the past two years, yet only 6\% report complete IaC coverage~\citep{firefly-state-iac-2025}. In such environments, ad hoc manual changes are difficult to audit, reproduce, and reconcile~\citep{nsync}, motivating IaC frameworks such as Terraform that express infrastructure through declarative, version-controlled specifications rather than imperative step-by-step operations~\citep{aws-iac,hashicorp-iac}.

\looseness=-1
However, handcrafting IaC configurations remains difficult because it requires substantial knowledge of cloud services, configuration semantics, and provider-specific tooling~\citep{llm-iac-survey}. 
This has fueled growing interest in using LLMs to generate IaC from natural language, as reflected in recent surveys and benchmarks studying cloud infrastructure generation and editing~\citep{llm-iac-survey,iaceval}.
However, generating correct IaC configurations is challenging, because it requires mapping underspecified user requests to complex, interdependent cloud configurations~\citep{iaceval}.

\begin{figure}[t]
    \centering
    \includegraphics[width=0.99\textwidth]{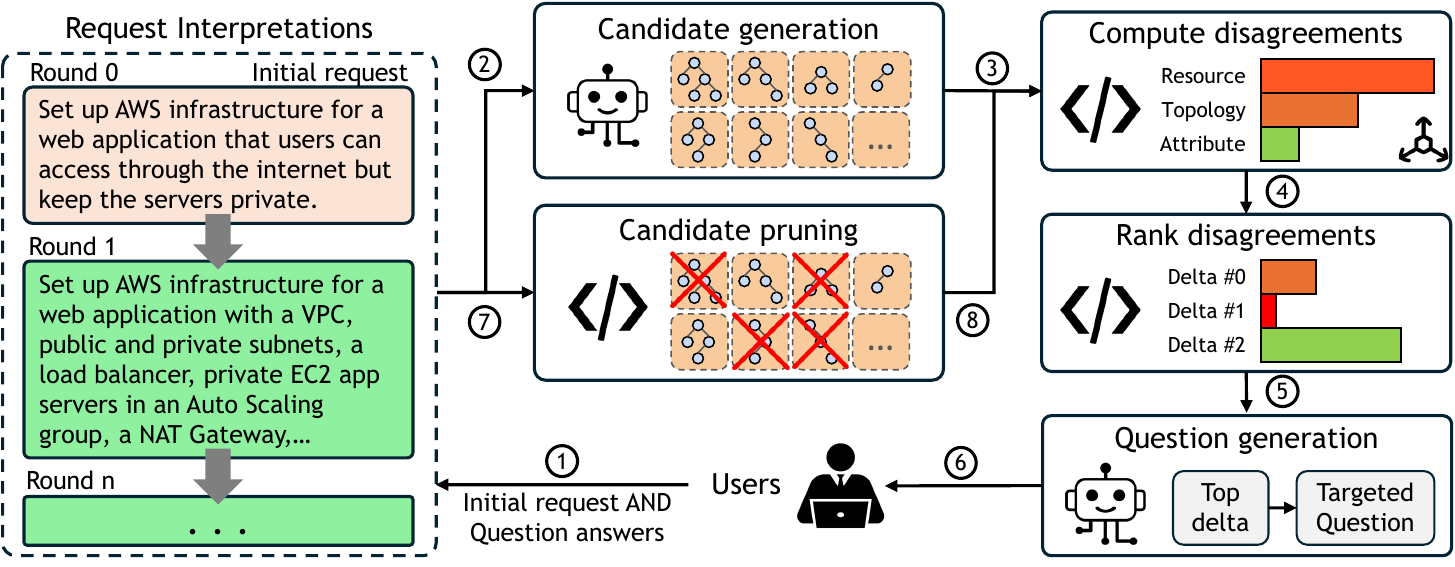}
    \caption{Overview of the iterative disambiguation process for interactive Infrastructure-as-Code synthesis.
\stepnum{1} Users provide an initial IaC request.
\stepnum{2} Initially, a pool of diverse structured specifications is generated based on the current interpretations.
\stepnum{3} Symbolically compute disagreements among the current pool of candidates.
\stepnum{4} These disagreements are ranked to identify the most informative differences measured by entropy.
\stepnum{5} Based on the top-ranked disagreement, the system generates a targeted clarification question.
\stepnum{6} Interaction with the user.
\stepnum{7} If the candidate pool is non-empty, prune candidates inconsistent with the user's answer.
\stepnum{8} The refined candidate pool is fed back into the next iteration, repeating the process until the interaction budget is exhausted. Otherwise, pool regeneration will be triggered.
}
    \label{fig:method}
\end{figure}

\section{Problem Formulation}
\label{sec:the-problem}

We study the problem of generating IaC configurations from ambiguous natural-language requests through interactive clarification.
A user's request expresses their \emph{explicit intent} $I$: the initial, underspecified prompt.
Because $I$ leaves many decisions open, it admits multiple plausible interpretations, which in turn induce a large space $\mathcal{C}(I)$ of distinct candidate configurations.
We assume that each request corresponds to a latent, intended target, represented in our setting by a reference configuration $c^\ast \in \mathcal{C}(I)$
that satisfies not only the explicit intent $I$ but also the user's \emph{implicit intent}, the unstated constraints and preferences.  
Since multiple syntactic realizations may represent the same design (e.g., different naming conventions), we normalize configurations to remove naming and representational differences. Architecturally distinct configurations, however, correspond to different interpretations and are therefore not treated as interchangeable.
The agent cannot observe the implicit intent, and thus cannot identify $c^\ast$ directly; it can only elicit information about it by asking the user a bounded number of clarification questions, progressively narrowing $\mathcal{C}(I)$ toward $c^\ast$.

\looseness=-1
\mypar{Structured specifications} 
Ambiguity in IaC synthesis arises along three axes: \textbf{resources}, \textbf{topology}, and \textbf{attributes}.
Together, these axes define the combinatorial space of infrastructure configurations.
IaC configurations consist of a set of resource blocks (see Figure~\ref{fig:intro-iac-code}), and we represent each configuration as a structured triple $s = (R, T, A)$, where $R$ is a multiset of resource types, $T \subseteq R \times R$ is a set of directed dependency edges (typed by their source and target resource types), and $A$ assigns to each resource a set of configuration key/value pairs. 
This decomposition induces a structured interface that supports more efficient and systematic intent elicitation by LLM-based agents.

\looseness=-1
\mypar{Interactive objective}
The agent is given a fixed budget of $K$ binary clarification questions.
At each round $t \in \{1,\dots,K\}$, it selects a question $q_t$, receives the user's answer $a_t \in \{\textsf{yes}, \textsf{no}\}$, and appends $(q_t, a_t)$ to the interaction history $H_t$.
After the budget is exhausted, the agent returns a single specification $\hat{c}$.
The goal is to make the final configuration as close as possible to the target $c^\ast$. Our evaluation measures this discrepancy using structural graph edit distance and exact resource-attribute matching.

\section{Method}
\label{sec:methods}

Since $c^\ast$ is unobserved and $\mathcal{C}(I)$ is intractably large, we approximate the configuration space with a finite \emph{candidate pool} $\mathcal{P}$ and use disagreement among candidates as a proxy for where the agent's uncertainty about $c^\ast$ concentrates. 
We model LLMs' alternative interpretations as a set of structured specifications, analyze the axes along which they disagree, and construct clarification questions based on the most informative disagreement. 
The agent iteratively asks one question per round, prunes the pool based on the user's answer, and regenerates candidates when the pool is exhausted.
After the budget is spent, the best surviving candidate is returned.

\mypar{Candidate generation} 
The candidate pool $\mathcal{P} = \{s_1,\dots,s_N\}$ serves as a concrete approximation of the space of plausible configurations $\mathcal{C}(I)$.
Given the user's request (and, in subsequent rounds, the accumulated question/answer history $H_t$), the agent generates $N$ candidate specifications via temperature sampling, so that they explore different resource selections, topology wiring, and attribute choices.
Candidates that share the same resource types and topology edges are deduplicated, as these higher-level axes determine the overall infrastructure design; attribute-level differences are still captured during disagreement detection.

\mypar{Computing disagreements} 
Because each candidate is a structured triple $(R,T,A)$, disagreements can be computed symbolically along each axis.
For resources, the agent extracts the set of resource types from each candidate and records any type that is present in some candidates but absent from others.
For topology, it extracts the set of directed dependency edges and similarly records edges on which candidates disagree.
For attributes, it groups candidates by resource type, then for each shared attribute key, checks whether candidates assign different values. 
Each disagreement partitions the pool into the candidates on either side of the difference. 

\mypar{Ranking disagreements} 
Not all disagreements are equally informative. 
A disagreement that has a vast majority carries little discriminative value, whereas an even split maximizes the information gained from a single binary question. 
We therefore rank disagreements by how evenly they divide the candidate pool, measured by the Shannon entropy of the induced split, and discard near-consensus disagreements that fall below an entropy floor.
To ensure balanced coverage, the surviving disagreements are ranked within each axis and then selected across axes in round-robin order, preventing any single axis from dominating the question budget. 

\mypar{Question generation and pruning}
In each round, the agent selects the top-ranked disagreement and converts it into a clarification question.
The disagreement already partitions the pool into two sides (e.g., candidates that include a NAT gateway vs.\ those that do not).
The agent passes it to an LLM, whose role is to rephrase the structural difference as a natural language yes/no question that a user can answer without knowledge of the underlying candidates.
Given the user's answer $a_t$, the agent retains only the candidates on the consistent side of the partition and discards the rest.
Because the partition is already known from the diff, pruning is immediate and deterministic: one answer can eliminate multiple candidates at once, efficiently narrowing the pool.

\mypar{Regeneration and termination} 
Regeneration is triggered when the pool is exhausted (fewer than two candidates remain).
The regeneration procedure repeats candidate construction, conditioning on both the original request and the full accumulated interaction history.
This lets the model produce candidates consistent with the information gathered so far while still exploring the remaining uncertainty; the structured representation makes newly generated candidates easy to deduplicate and filter against prior interactions, avoiding revisiting resolved disagreements.
The interaction terminates when the budget $K$ is exhausted or when the agent can no longer produce structurally distinct candidates, indicating that its uncertainty has been sufficiently resolved.
Upon termination, the agent returns the best surviving candidate; if none survives, it performs one final generation conditioned on the original request and the full interaction history.

%% file: evaluation6.tex

\section{Experimental Setup}
\label{sec:eval-setup}

\begin{table}[tb]
\centering
\small
\input{tables/ab/human_validation}
\caption{Human validation of \textsc{Ambig-IaC} and IaC-Eval on 100 samples, with normalized ratings and inter-coder agreement measured by Gwet’s AC. Higher ratings indicate greater naturalness, ambiguity, and target consistency.
}
\label{tab:human-val}
\end{table}

\textbf{Dataset.}
We construct \textsc{Ambig-IaC}, a benchmark of 300 Terraform tasks sourced from IaC-Eval~\citep{iaceval}.
We manually verified and fixed the reference programs using \texttt{terraform plan}, filtering out programs that could not be validated.
The dataset covers 167 unique AWS resource types across 44 service families (e.g., VPC, S3, IAM, RDS, Lambda).
To introduce realistic ambiguity, we use an LLM to rewrite each original task prompt into a high-level, goal-oriented request that preserves consistency with the reference configuration but admits multiple plausible architectures.
For example, ``Configure a weighted routing policy that splits users between two db\_instances that are replicas of a main db\_instance'' becomes ``Our database is getting slow for users in other regions---need to fix that and make reads faster globally.''
A manual inspection of 100 sampled instances (performed by three domain experts) confirms that the rewritten prompts are natural yet answerable tasks: relative to IaC-Eval, \textsc{Ambig-IaC} scores substantially higher on Naturalness and Ambiguity while preserving Target consistency, with substantial-to-almost-perfect inter-coder agreement (Table~\ref{tab:human-val}).

\textbf{User Proxy.}
We simulate user responses with an LLM-based oracle, following prior work~\citep{tomswe, clarifygpt, Sanidhya-2025-interact, kobalczyk2025atd, zhang2025modelingfutureconversationturns}.
Given a ground-truth Terraform configuration and a binary clarification question, the proxy answers based solely on whether the reference configuration implies yes or no, and is explicitly instructed to answer strictly from the reference, with no external knowledge or assumptions.
All methods interact with the same oracle under the same question budget, so the information channel is held fixed and differences between methods reflect question quality alone.

\textbf{Baselines.}
Under the same question budget~$K$, we compare against four interactive baselines:
\begin{itemize}
  \item \textbf{Direct Question Generation}: the LLM directly generates clarification questions.
  \item \textbf{Best-of-N}: at each round the LLM generates $N$ candidate questions conditioned on prior Q\&A, and a separate ranker LLM selects one.
  \item \textbf{Self-Consistency}: $N$ candidate questions are generated, embedded, and clustered; the question closest to the centroid of the largest cluster is selected, favoring questions that the model consistently considers important.
  \item \textbf{Active Task Disambiguation (ATD)}~\citep{kobalczyk2025atd}: ranks candidate questions by expected information gain, estimated over a grid of candidate questions $\times$ solutions; evaluated at $K\in\{5,10\}$ due to high costs.
\end{itemize}
All baselines share the same final generation step: after $K$ rounds of Q\&A, the accumulated interaction history is used to construct a clarified intent, from which the final specification is generated.

\textbf{Metrics.}
We evaluate generated specifications against the ground-truth reference along the two axes of IaC ambiguity: whether the infrastructure has the right \emph{structure}, and whether its concrete \emph{values} are exactly right. Both metrics operate on specs normalized to a common label namespace.
\textbf{Structure} (Struct.) scores graph similarity via the Graph Edit Distance (GED) between the reference and generated resource graphs, using unit costs for node/edge insertions, deletions, and cross-type substitutions. It jointly penalizes missing or extra resources \emph{and} incorrect wiring; the distance is normalized by graph size and inverted, so $1.0$ is an exact structural match.
\textbf{Res+Attr F1} (R+A F1) is a deliberately strict, exact-match test of concrete configuration values.
Each spec is reduced to an occurrence-aware multiset of resource instances and attribute key--value pairs (multiplicity counts: three \texttt{aws\_subnet} in the reference but two generated yields at most two matches), and the score is the exact-match F1 between the two multisets after canonicalizing booleans, numbers, and letter case.
We grant no embedding or fuzzy credit, because values that are close in form can be incompatible in behavior (\texttt{10.0.0.0/16} vs.\ \texttt{10.1.0.0/16}); free-form naming keys and cross-resource references (already scored by \emph{Structure}) are excluded.
All methods are evaluated against the same reference configurations under the same interaction protocol, so comparisons are made under a consistent target and scoring procedure.

\section{Results and Analysis}
\label{sec:results}

Our evaluation centers on four research questions:
\begin{itemize}
  \item \textbf{RQ1:} Does structure-aware disambiguation outperform structure-agnostic baselines?
  \item \textbf{RQ2:} Does structure-aware disambiguation performance scale with the number of rounds?
  \item \textbf{RQ3:} Does the method generalize across different LLMs?
  \item \textbf{RQ4:} Are the reported gains statistically robust?
\end{itemize}
Following the main results, we present ablation studies on hyperparameter sensitivity, cross-axes balancing, and deployment-time validity.


\begin{table}[t]
\centering
\small
\resizebox{\textwidth}{!}{\input{tables/ab/main_results_v6}}
\caption{Main results on \textsc{Ambig-IaC} using GPT-4o-mini. We report means with 95\% task-level bootstrap CIs over 300 tasks (10,000 resamples). Struct. is normalized-GED similarity; R+A F1 is exact-match resource-and-attribute F1; and $\Delta$ gives each baseline's absolute and relative difference from ours. ATD is evaluated at $K\in\{5,10\}$ due to its substantially higher cost.}
\label{tab:main}
\end{table}

\subsection{RQ1: Overall Performance}
\label{subsec:rq1}

Structure-aware disambiguation outperforms all structure-agnostic baselines: across budgets of $K{=}5$, $10$, and $15$, our method achieves the best Structure score at every budget and the best R+A F1 at the two larger budgets; its lead over the strongest baseline widens as the budget grows (Table~\ref{tab:main}).
We observe that test-time inference techniques (Best-of-N, Self-Consistency) do not offer a meaningful improvement over direct question generation, and Self-Consistency is the weakest method overall: question consensus does not align with informativeness in a highly structured configuration space, where useful questions are those that target the specific points on which interpretations diverge.
ATD, also an information-gain method, generates candidate questions and scores them over a question~$\times$~solution grid rather than deriving them from computed spec deltas.
It trails our Structure score while spending $4\times$ our tokens and $11\times$ our wall-clock per task (detailed cost results are in the supplementary material), and its one nominal edge, R+A F1 at $K{=}5$, sits well inside our confidence interval (RQ4).
For controlled comparison, all methods use the same binary-question interface, which restricts the information conveyed per turn. 
Richer response formats and persistent user-preference models may improve absolute performance, but evaluating these extensions requires interaction data that is not currently available for IaC.
\emph{\textbf{Finding 1:} LLMs do not natively ask informative domain-specific clarification questions; deriving questions from computed structural disagreement is simultaneously more informative and dramatically cheaper — matching or exceeding ATD while using 4$\times$ fewer tokens and 11$\times$ less wall-clock time per task.}


\subsection{RQ2: Interaction Scaling}

\begin{table}[t]
\centering
\small
\input{tables/ab/rounds_gain_v6}
\caption{Question-budget scaling (GPT-4o-mini): average gain in points per 5-question budget increase, with the relative gain in parentheses ($5{\to}10$ and $10{\to}15$ increments, from Table~\ref{tab:main}).}
\label{tab:rounds-gain}
\end{table}

\begin{figure}[t]
\centering
\includegraphics[width=0.8\columnwidth]{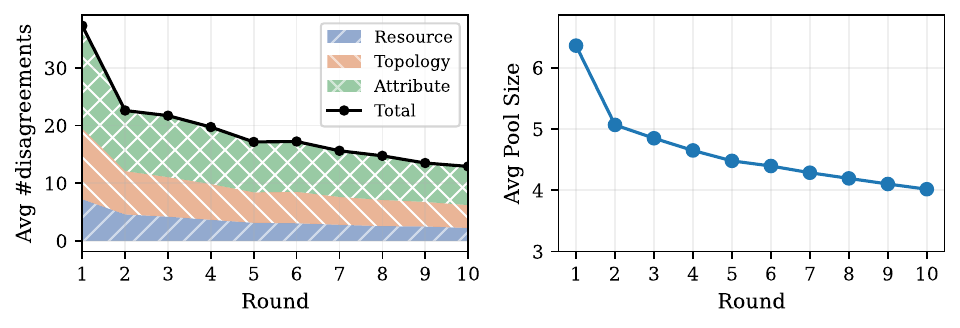}
\caption{Per-round dynamics show the candidates converging toward a shared interpretation.}
\label{fig:round-dynamics}
\end{figure}

\begin{figure}[t]
\centering
\includegraphics[width=0.8\columnwidth]{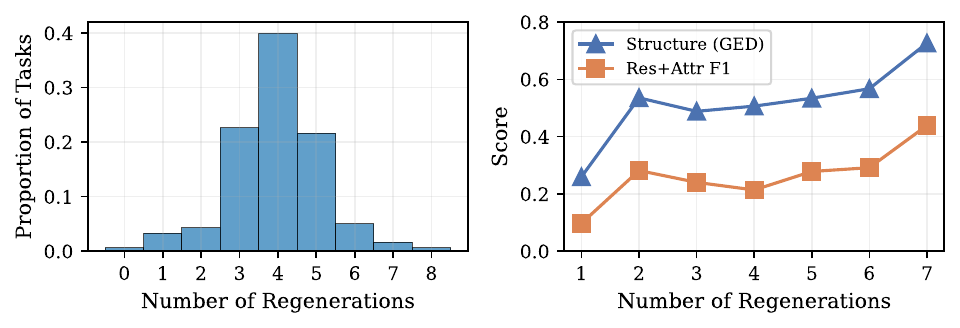}
\caption{Most tasks require three to five regenerations, with performance generally improving as regenerations increase.}
\label{fig:regen-dynamics}
\end{figure}

Most methods benefit from additional questions, as shown in Table~\ref{tab:rounds-gain}, with ours gaining the most on both metrics. 
Self-Consistency even shows a slight decrease, suggesting that consensus-based selection saturates quickly. The widening gap indicates that our method makes more effective use of
each additional question. Baselines generate questions without explicit awareness of which axes remain ambiguous, so later questions are more
likely to be redundant or low-value. 
In contrast, our disagreement-driven approach continually re-estimates where ambiguity lies after each answer, directing subsequent questions to the most informative remaining differences.
Figure~\ref{fig:round-dynamics} shows that disagreement counts decrease steadily across all three axes (left), and resource-level disagreements are resolved first, then topology, then attributes, 
while the pool size shrinks in step (right), converging toward a shared interpretation.
Figure~\ref{fig:regen-dynamics} examines regeneration: most tasks require 3--5 regenerations (left), indicating that the candidate pool is typically exhausted within 2--3 questions, as each question is sufficiently discriminative to eliminate a substantial fraction of candidates.
We see a positive correlation between regeneration frequency and performance (right).
This suggests that conditioned regeneration---producing new candidates consistent with accumulated Q\&A history---is an effective mechanism for progressive refinement, as each cycle introduces candidates better aligned with the user's intent.
\emph{\textbf{Finding 2:} Additional interaction helps the most when the agent explicitly tracks unresolved ambiguity.}

\begin{table}[t]
\centering
\small
\begin{minipage}[t]{0.545\textwidth}
\vspace{0pt}
\centering
\setlength{\tabcolsep}{4pt}
\resizebox{\linewidth}{!}{\input{tables/ab/models_res_v5}}
\caption{Generalization across five backbone LLMs ($K{=}5$). Our method is best on both metrics for every model.}
\label{tab:models}
\end{minipage}\hfill
\begin{minipage}[t]{0.44\textwidth}
\vspace{0pt}
\centering
\resizebox{\linewidth}{!}{\input{tables/ab/seeded_runs_v5}}
\caption{Seeded runs (Claude Opus 4.6, $K{=}5$): mean\,$\pm$\,std over three seeds. 
}
\label{tab:seeded}

\vspace{14pt}
\centering
\input{tables/ab/hyperparam_v5}
\caption{Hyperparameter sweep over pool size $N$ and temperature $T$ (Claude Opus 4.6, $K{=}5$).
}
\label{tab:hyperparam}
\end{minipage}
\end{table}

\subsection{RQ3: Cross-Model Generalization}
The gains generalize across backbones: we re-ran the full pipeline on five models spanning open and closed families, and our method achieves the best results on both metrics for every one of them (Table~\ref{tab:models}).
Stronger backbones lift all methods, but the relative ordering remains unchanged; the lead is largest on Qwen3-Coder and remains clear on the strongest backbone, Claude Opus 4.6.


\subsection{RQ4: Statistical Robustness}
The reported gains are statistically robust to both sampling stochasticity and task selection, which we assess in two ways.
Across three seeded trials on the strongest backbone (Table~\ref{tab:seeded}), our lead over the strongest baseline is many times the run-to-run standard deviation on both metrics.
The task-level bootstrap CIs in Table~\ref{tab:main} tell the same story at the population level: on Structure, our interval is disjoint from every baseline's at $K{=}10$ and $K{=}15$, and essentially disjoint at $K{=}5$; on the stricter R+A F1, our mean lies above every baseline's upper bound at $K{=}10$ and $K{=}15$.

\subsection{Ablation Studies}
\label{subsec:ablation}

\begin{table}[t]
\centering
\small
\setlength{\tabcolsep}{4pt}
\input{tables/ab/crossdim_v5}
\caption{Cross-axes balancing ablation (GPT-4o-mini). 
}
\label{tab:crossdim}
\end{table}

\mypar{Hyperparameter sensitivity}
We now vary the pool size $N{\in}\{5,10,20\}$ and sampling temperature $T{\in}\{0.5,0.7,0.9\}$ (Table~\ref{tab:hyperparam}).
The spread across all nine settings is small; neither knob shows a monotonic trend, and every cell outperforms the strongest same-backbone baseline from Table~\ref{tab:models}.
Notably, the paper default ($N{=}10$, $T{=}0.7$) is not even the strongest cell in the sweep, yet it wins everywhere it is compared: the gains come from the disagreement-driven procedure, not from a fragile hyperparameter choice.

\mypar{Cross-axes balancing}
Our method allocates questions round-robin across the resource, topology, and attribute axes rather than by raw entropy alone.
We see a performance drop as shown in Table~\ref{tab:crossdim}, suggesting that the round-robin balancing does help with IaC disambiguation. 

\mypar{Deployability}
Finally, we verify that generated specifications pass Terraform's own validation toolchain.
We convert each specification into a program \texttt{main.tf} using an LLM, and run the standard pipeline (\texttt{init}\,$\rightarrow$\,\texttt{validate}\,$\rightarrow$\,\texttt{plan}): $90\%$ of tasks pass on one-shot emission and $97$--$99\%$ after a single error-guided repair pass, for every method.
What separates methods is whether the \emph{right} infrastructure is specified: the resources, wiring, and values the user actually intended, which is exactly what our metrics measure and where our method leads.
\emph{\textbf{Finding 3:} LLMs can generate runnable IaC, but deployment success does not guarantee alignment with user intent; mis-specified configurations may deploy successfully despite unresolved ambiguity.}

%% file: tables/ab/human_validation.tex
\begin{tabular}{l ccc}
\toprule
\textbf{Dataset} & \textbf{Natural.} & \textbf{Ambig.} & \textbf{Target Cons.} \\
\midrule
\multicolumn{4}{l}{\emph{Normalized score} ($0$--$1$)} \\
\addlinespace[1pt]
IaC-Eval  & 0.25 & 0.24 & 0.90 \\
\graycell \textsc{Ambig-IaC} & \graycell \textbf{0.78} & \graycell \textbf{0.89} & \graycell \textbf{0.99} \\
\midrule
\multicolumn{4}{l}{\emph{Inter-coder agreement} (Gwet's AC)} \\
\addlinespace[1pt]
IaC-Eval  & 0.71 & 0.82 & 0.70 \\
\textsc{Ambig-IaC} & 0.65 & 0.82 & 0.98 \\
\bottomrule
\end{tabular}

%% file: tables/ab/main_results_v6.tex
\begin{tabular}{l|c|cc|cc}
  \toprule
  \textbf{Method} & \textbf{Budget $K$} & \textbf{Struct.\ (\%)} & \textbf{$\Delta$ Struct.} & \textbf{R+A F1 (\%)} & \textbf{$\Delta$ F1} \\
  \midrule
  Direct Q Generation & 5 & 43.08 \ci{39.90}{46.31} & -5.87 (\dnar{11.99\%}) & 17.69 \ci{15.40}{20.05} & -4.35 (\dnar{19.74\%}) \\
  Best-of-N & 5 & 41.28 \ci{37.97}{44.68} & -7.67 (\dnar{15.67\%}) & 18.18 \ci{15.80}{20.68} & -3.86 (\dnar{17.51\%}) \\
  Self-Consistency & 5 & 38.59 \ci{35.48}{41.73} & -10.36 (\dnar{21.16\%}) & 15.63 \ci{13.40}{17.97} & -6.41 (\dnar{29.08\%}) \\
  ATD & 5 & 42.23 \ci{38.55}{45.85} & -6.72 (\dnar{13.73\%}) & \textbf{22.70} \ci{20.21}{25.21} & +0.66 (\upar{2.99\%}) \\
  \graycell \textbf{Ours} & \graycell 5 & \graycell \textbf{48.95} \ci{45.95}{51.91} & \graycell -- & \graycell 22.04 \ci{19.57}{24.63} & \graycell -- \\
  \midrule
  Direct Q Generation & 10 & 44.38 \ci{41.31}{47.52} & -6.60 (\dnar{12.95\%}) & 18.72 \ci{16.58}{20.97} & -5.49 (\dnar{22.68\%}) \\
  Best-of-N & 10 & 43.52 \ci{40.24}{46.79} & -7.46 (\dnar{14.63\%}) & 18.91 \ci{16.48}{21.34} & -5.30 (\dnar{21.89\%}) \\
  Self-Consistency & 10 & 38.41 \ci{35.27}{41.71} & -12.57 (\dnar{24.66\%}) & 16.30 \ci{14.04}{18.66} & -7.91 (\dnar{32.67\%}) \\
  ATD & 10 & 41.89 \ci{37.95}{45.83} & -9.09 (\dnar{17.83\%}) & 24.01 \ci{21.38}{26.69} & -0.20 (\dnar{0.83\%}) \\
  \graycell \textbf{Ours} & \graycell 10 & \graycell \textbf{50.98} \ci{48.02}{53.98} & \graycell -- & \graycell \textbf{24.21} \ci{21.62}{26.87} & \graycell -- \\
  \midrule
  Direct Q Generation & 15 & 46.31 \ci{43.28}{49.42} & -8.53 (\dnar{15.55\%}) & 18.45 \ci{16.33}{20.62} & -8.08 (\dnar{30.46\%}) \\
  Best-of-N & 15 & 43.12 \ci{39.86}{46.48} & -11.72 (\dnar{21.37\%}) & 19.52 \ci{17.08}{22.04} & -7.01 (\dnar{26.42\%}) \\
  Self-Consistency & 15 & 38.19 \ci{34.95}{41.39} & -16.65 (\dnar{30.36\%}) & 16.12 \ci{13.81}{18.50} & -10.41 (\dnar{39.24\%}) \\
  \graycell \textbf{Ours} & \graycell 15 & \graycell \textbf{54.84} \ci{51.87}{57.86} & \graycell -- & \graycell \textbf{26.53} \ci{23.88}{29.28} & \graycell -- \\
  \bottomrule
\end{tabular}

%% file: tables/ab/rounds_gain_v6.tex
\begin{tabular}{l cc}
\toprule
\textbf{Method} & \textbf{$\Delta$ Struct.} & \textbf{$\Delta$ R+A F1} \\
\midrule
  Direct Q Gen.       & +1.62 (\upar{3.68\%}) & +0.38 (\upar{2.19\%}) \\
  Best-of-N           & +0.92 (\upar{2.25\%}) & +0.67 (\upar{3.62\%}) \\
  Self-Consistency    & -0.20 (\dnar{0.52\%}) & +0.25 (\upar{1.59\%}) \\
  \graycell \textbf{Ours} & \graycell \textbf{+2.95 (\upar{5.86\%})} & \graycell \textbf{+2.25 (\upar{9.71\%})} \\
\bottomrule
\end{tabular}

%% file: tables/ab/models_res_v5.tex
\begin{tabular}{ll cc}
\toprule
\textbf{Model} & \textbf{Method} & \textbf{Struct.} & \textbf{R+A F1} \\
\midrule
  \multirow{4}{*}{Claude Opus 4.6}
   & Direct Q Gen.    & 53.17 & 26.39 \\
   & Best-of-N        & 57.67 & 30.07 \\
   & Self-Consistency & 56.95 & 28.87 \\
   & \graycell \textbf{Ours} & \graycell \textbf{64.69} & \graycell \textbf{32.31} \\
\midrule
  \multirow{4}{*}{GPT-4.1-mini}
   & Direct Q Gen.    & 50.77 & 22.42 \\
   & Best-of-N        & 53.06 & 23.69 \\
   & Self-Consistency & 52.58 & 22.26 \\
   & \graycell \textbf{Ours} & \graycell \textbf{57.33} & \graycell \textbf{26.27} \\
\midrule
  \multirow{4}{*}{Qwen3-Coder NEXT}
   & Direct Q Gen.    & 49.05 & 19.62 \\
   & Best-of-N        & 48.17 & 20.28 \\
   & Self-Consistency & 47.66 & 18.40 \\
   & \graycell \textbf{Ours} & \graycell \textbf{58.53} & \graycell \textbf{25.19} \\
\midrule
  \multirow{4}{*}{Llama-4-Scout}
   & Direct Q Gen.    & 48.92 & 17.37 \\
   & Best-of-N        & 50.29 & 18.14 \\
   & Self-Consistency & 49.42 & 17.15 \\
   & \graycell \textbf{Ours} & \graycell \textbf{54.02} & \graycell \textbf{21.23} \\
\midrule
  \multirow{4}{*}{GPT-4o-mini}
   & Direct Q Gen.    & 43.08 & 17.69 \\
   & Best-of-N        & 41.28 & 18.18 \\
   & Self-Consistency & 38.59 & 15.63 \\
   & \graycell \textbf{Ours} & \graycell \textbf{48.95} & \graycell \textbf{22.04} \\
\bottomrule
\end{tabular}

%% file: tables/ab/seeded_runs_v5.tex
\begin{tabular}{lcc}
\toprule
\textbf{Method} & \textbf{Struct.} & \textbf{R+A F1} \\
\midrule
  Direct Q Generation & 53.46\stdd{0.37} & 25.75\stdd{0.60} \\
  Best-of-N           & 57.91\stdd{0.24} & 30.37\stdd{0.83} \\
  Self-Consistency    & 57.46\stdd{1.25} & 28.83\stdd{0.51} \\
  \graycell \textbf{Ours} & \graycell \textbf{63.95}\stdd{0.68} & \graycell \textbf{31.95}\stdd{0.32} \\
\bottomrule
\end{tabular}

%% file: tables/ab/hyperparam_v5.tex
\begin{tabular}{cc|cc}
\toprule
\textbf{$N$} & \textbf{$T$} & \textbf{Struct.} & \textbf{R+A F1} \\
\midrule
  5  & 0.5 & 67.27 & 33.03 \\
  5  & 0.7 & 63.19 & 33.11 \\
  5  & 0.9 & 62.24 & 32.56 \\
  \midrule
  10 & 0.5 & 63.62 & 32.85 \\
  \graycell 10 & \graycell 0.7$^{\dagger}$ & \graycell 64.69 & \graycell 32.31 \\
  10 & 0.9 & 66.10 & 33.31 \\
  \midrule
  20 & 0.5 & 63.69 & 32.81 \\
  20 & 0.7 & 64.27 & 33.14 \\
  20 & 0.9 & 65.49 & 34.66 \\
\bottomrule
\end{tabular}

%% file: tables/ab/crossdim_v5.tex
\begin{tabular}{c|cc|cc}
  \toprule
  \textbf{$K$} & \textbf{Struct.} & \textbf{$\Delta$ Struct.} & \textbf{R+A F1} & \textbf{$\Delta$ F1} \\
  \midrule
  5  & 47.18 & $-$1.77 (\dnar{3.62\%}) & 21.08 & $-$0.96 (\dnar{4.36\%}) \\
  10 & 49.74 & $-$1.24 (\dnar{2.43\%}) & 23.25 & $-$0.96 (\dnar{3.97\%}) \\
  15 & 51.65 & $-$3.19 (\dnar{5.82\%}) & 24.34 & $-$2.19 (\dnar{8.25\%}) \\
  \bottomrule
\end{tabular}

%% file: related_work2.tex
\section{Related Work}
\label{sec:related-work}




\textbf{Agents for cloud management.} Recent work has explored LLM-based systems for cloud management, including IaC generation benchmarks~\citep{iaceval}, agentic systems for autonomous cloud operations and incident analysis~\citep{shetty2024building,AIOpsLab_mlsys25,RCACopilot_eurosys24,nsync, terrafault, lilac}, provider-integrated assistants such as Azure Copilot, Gemini for Google Cloud, and Amazon Q~\citep{azure-copilot,gcp-gemini,aws-q}, and broader discussions of cloud AI agents~\citep{cloudagent_vision}. 
These efforts demonstrate the promise of LLM agents for IaC generation, diagnosis, and cloud operational assistance. 
However, they generally assume that the given tasks are well specified and do not study how an agent should resolve ambiguous user intent before acting.

\textbf{Active task disambiguation.} 
Recent work on active task disambiguation selects clarification questions by estimating how candidate answers partition a space of plausible solutions~\citep{kobalczyk2025atd}. 
ATD ranks freely generated candidate questions by estimated information gain over a question $\times$ solution grid; we instead derive questions directly from symbolically computed disagreements across resources, topology, and attributes, which is both substantially cheaper (Section~\ref{subsec:rq1}) and better targeted. Our method further coordinates resolution across rounds via cross-dimension balancing rather than ranking by aggregate uncertainty alone.
This structured selection strategy produces better final specifications than entropy-only ranking (Section~\ref{subsec:ablation}).

\looseness=-1
\textbf{Clarification via fine-tuning.} 
Prior work learns clarification behavior in bounded QA, preference-elicitation, and software-engineering settings through preference optimization, self-training, multi-turn policy learning, or persistent user modeling~\citep{ambigqa,zhang2025modelingfutureconversationturns,stargate,l2c,tomswe}. These approaches generally assume relatively stable task distributions and rely on fine-tuning, synthetic supervision, or learned user models. 
In contrast, cloud IaC is a fast-evolving domain where provider APIs and specifications change rapidly, high-quality training data is scarce, and ambiguity is structured across resources, topology, and attributes.

\textbf{Clarification in code.} 
The closest engineering analogues are code-oriented systems that resolve ambiguity by comparing competing realizations of the user's intent and querying the user on distinguishing evidence~\citep{clarifygpt,ITDCG}. Recent work in software-engineering agents further shows that interaction improves performance on underspecified tasks~\citep{Sanidhya-2025-interact}. 
These results demonstrate the value of clarification in engineering domains, but all rely on executable artifacts, generated tests, or cheap iterative repair — feedback unavailable in IaC, where execution provisions real infrastructure and ambiguity must be resolved before deployment.

%% file: conclusion.tex
\section{Conclusion and Future Work}
\label{sec:conclusion}


We have presented a disagreement-driven disambiguation framework for interactive IaC synthesis that leverages configuration structure for multi-turn clarification. We have also introduced \textsc{Ambig-IaC}, a 300-task, expert-verified benchmark of IaC tasks with ambiguous prompts. Our method outperforms structure-agnostic baselines, with gains that scale with the interaction budget and generalize across models. 

Our evaluation relies on a simulated protocol: prompts rewritten from verified references and answers from a reference-grounded oracle, following standard practice; expert validation supports prompt realism (Table~\ref{tab:human-val}), and since all methods face the same oracle, relative comparisons remain fair. Validating these findings with real users is important future work. 
More broadly, beyond IaC, this work suggests that disagreement-driven clarification may provide a general approach for resolving ambiguity in other structured synthesis tasks.

%% file: appendix.tex
\noindent
This supplement contains: (i)~pseudocode and sub-routine definitions for the
multi-level disambiguation algorithm (Section~\ref{app:algo}); (ii)~the
experiment configuration---models, roles, decoding settings, and method
hyperparameters (Section~\ref{app:config}); (iii)~formal definitions of the two
evaluation metrics, the GED-based Structure score and the exact-match Res+Attr
F1, together with the label normalization both share
(Section~\ref{app:metrics}); (iv)~per-task token and wall-clock costs for all
methods, models, and interaction budgets (Section~\ref{app:cost}); (v)~an
ablation of the output protocol, separating the effect of question selection
from that of final-spec emission (Section~\ref{app:output}); and (vi)~the
system and user prompts used at each stage of the pipeline
(the \emph{Prompts} section at the end of this supplement).

\section{Multi-Level Disambiguation Algorithm}
\label{app:algo}

Algorithm~\ref{alg:mld} gives the top-level loop of our method;
the subsections below define each sub-routine.

\subsection{Notation}
\label{app:notation}

Each candidate configuration is a structured triple
\begin{equation}
  s = (R,\,T,\,A),
  \label{eq:spec-triple}
\end{equation}
where $R$ is a multiset of resource types, $T \subseteq R \times R$ is a set of
directed dependency edges, typed by their source and target resource types, and
$A$ assigns each resource a set of configuration key--value pairs.
Concretely, a resource is a label$\,\mapsto\,$Terraform-address entry, and
$\rtype(a)$ denotes the type extracted from address $a$, e.g.,
$\rtype(\code{aws\_vpc.main})=\code{aws\_vpc}$. The agent maintains a
\emph{candidate pool} $\mathcal{P}=\{s_1,s_2,\dots\}$ approximating the space of
plausible configurations, and an \emph{interaction history} $H$ of
question--answer pairs. The agent may ask at most $K$ binary questions; $N$ is
the pool size and $\theta$ the entropy floor.

\begin{algorithm}[t]
\caption{Multi-Level Disambiguation}
\label{alg:mld}
\textbf{Input}: request $I$; budget $K$; pool size $N$; entropy floor $\theta$\\
\textbf{Output}: final specification $\hat{c}$
\begin{algorithmic}[1]
\STATE $H \gets \emptyset$
\STATE $\mathcal{P} \gets \textsc{Generate}(I, H, N)$
\FOR{$t = 1$ \TO $K$}
    \IF{$|\mathcal{P}| < 2$}
        \STATE $\mathcal{P} \gets \mathcal{P} \cup \textsc{Generate}(I, H, N)$ 
    \ENDIF
    \STATE $\mathcal{D} \gets \textsc{Disagreements}(\mathcal{P})$
    \IF{$\mathcal{D} = \emptyset$}
        \STATE \textbf{break} 
    \ENDIF
    \STATE $\mathcal{D}_{>\theta} \gets \{\, f \in \mathcal{D} : H_{\mathrm{ent}}(f) > \theta \,\}$
    \IF{$\mathcal{D}_{>\theta} \neq \emptyset$}
        \STATE $\mathcal{D} \gets \mathcal{D}_{>\theta}$ \hfill\COMMENT{drop near-consensus splits}
    \ENDIF
    \STATE $F^\ast \gets \textsc{RoundRobinShortlist}(\mathcal{D})$ \hfill\COMMENT{$\le 5$ diffs, interleaved across axes}
    \STATE $(q_t,\, \mathcal{P}^{\textsf{yes}},\, \mathcal{P}^{\textsf{no}}) \gets \textsc{Verbalize}(F^\ast)$
    \STATE ask $q_t$; observe $a_t \in \{\textsf{yes},\textsf{no}\}$
    \STATE $\mathcal{P} \gets \mathcal{P} \cap \mathcal{P}^{a_t}$ \hfill\COMMENT{deterministic prune}
    \STATE $H \gets H \cup \{(q_t, a_t)\}$
\ENDFOR
\STATE \textbf{return} $\textsc{Select}(\mathcal{P}, I, H)$
\end{algorithmic}
\end{algorithm}

\subsection{Candidate Generation}
\label{app:gen}

\textsc{Generate}$(I, H, N)$ produces $N$ candidates conditioned on the request
$I$ and the history $H$. One \emph{anchor} spec is generated at temperature~$0$
under a plain generation prompt with no diversity instructions, so the pool
always contains a canonical interpretation. The remaining $N{-}1$ specs are sampled at a higher temperature ($0.7$ by default).

\subsection{Disagreement Computation}
\label{app:diff}

\textsc{Disagreements}$(\mathcal{P})$ compares the candidates axis by axis;
each disagreement $f$ partitions the candidates it applies to.

\begin{itemize}[leftmargin=1.4em,nosep]
  \item \textbf{Resources.} For each type $r$ occurring in any candidate, let
        $\mathrm{present}(r)=\{i : r \in R(s_i)\}$ and
        $\mathrm{absent}(r)=\{1,\dots,|\mathcal{P}|\}\setminus\mathrm{present}(r)$.
        A disagreement is recorded when both sets are non-empty.
  \item \textbf{Topology.} The same test, per type-level directed edge:
        $\mathrm{present}(e)=\{i : e \in T(s_i)\}$.
  \item \textbf{Attributes.} Candidates are grouped by resource type. For each
        type $r$ and each key $k$ set on $r$ by any candidate in the group,
        every candidate containing $r$ contributes a value: the value it
        assigns, or \emph{unset} if it does not set $k$. A disagreement is
        recorded when these values are not all identical (values are compared
        after canonical JSON serialization).
\end{itemize}

\subsection{Entropy Ranking and Cross-Dimension Balancing}
\label{app:rank}

A near-unanimous split carries little information; an even split is worth a
question. Let $\{\mathcal{P}_b\}$ be the parts into which $f$ divides the
candidates it applies to---present vs.\ absent for resources and topology, one
part per distinct value (including \emph{unset}) for attributes. We score $f$
by the Shannon entropy of this split,
\begin{equation}
  H_{\mathrm{ent}}(f) = -\sum_{b} \rho_b \log_2 \rho_b,
  \qquad \rho_b = \frac{|\mathcal{P}_b|}{\sum_{b'} |\mathcal{P}_{b'}|}.
  \label{eq:entropy}
\end{equation}
Disagreements with $H_{\mathrm{ent}} \le \theta$ ($0.5$ in all experiments)
are discarded as near-consensus splits; if the floor would discard every
disagreement while the candidates still differ, the unfiltered set is used
instead, so a round is never wasted. Within each axis, the survivors are
ranked by descending entropy, and \textsc{RoundRobinShortlist} interleaves
the per-axis rankings in round-robin order, keeping at most five: the
shortlist $F^\ast$ thus mixes the most informative disagreements from all
three axes. This balancing keeps any single axis from monopolizing the
question budget; the main paper's ablation shows it outperforms ranking by
entropy alone.

\subsection{Question Generation and Pruning}
\label{app:qgen}

\textsc{Verbalize}$(F^\ast)$ renders the shortlist as a structural diff
summary---each disagreement listed with the candidate indices on either
side---preceded by a short resource/topology synopsis of every candidate,
and passes it to an LLM (the exact prompt and summary format are in
the \emph{Prompts} section). The LLM emits \emph{one} yes/no question that
targets the most informative ambiguity in the summary, phrased so a user can
answer it without seeing the candidates (e.g., a resource disagreement on
\code{aws\_nat\_gateway} becomes ``Should the infrastructure include a NAT
gateway?''); the lower-ranked shortlist entries serve as context, helping
the LLM keep the question precise and non-overlapping. 
Pruning requires no further LLM call and is deterministically answered.
One answer can eliminate several candidates at once, and survivors are kept
exactly as generated, so no editing errors accumulate. If the observed
answer's keep set is empty or inconsistent with every surviving candidate,
the pool is discarded, which triggers regeneration at the start of the next round.




\section{Experiment Configuration}
\label{app:config}

\paragraph{Roles.}
Every run is driven by an LLM configuration with three roles: the
\emph{generator} (candidate and final spec generation), the \emph{clarifier}
(question generation), and the \emph{proxy} (the simulated user). The clarifier
and generator use the backbone under evaluation; the proxy is GPT-4o-mini at
temperature~$0$ in every experiment, so all methods and backbones face the same
answer channel.

\paragraph{Backbones.}
We evaluate GPT-4o-mini, GPT-4.1-mini, Claude Opus 4.6,
Llama-4-Scout-17B-16E-Instruct, and
Qwen3-Coder NEXT, each accessed through its provider's standard API. The
pipeline's prompts are reproduced in the \emph{Prompts} section at the end of
this supplement.


\paragraph{Method hyperparameters.}
Table~\ref{tab:hyperparams} lists the settings of our method; all runs use the
values shown unless noted otherwise. The method has three design
knobs: the pool size $N$, the diverse-sampling temperature
$T_{\mathrm{div}}$, and the entropy floor $\theta$. The main paper's
sensitivity ablation sweeps the two knobs, $N\in\{5,10,20\}$ and
$T_{\mathrm{div}}\in\{0.5,0.7,0.9\}$, on a 100-task sample (Claude Opus 4.6,
$K{=}5$), with $\theta$ held fixed.

\begin{table}[t]
\centering
\small
\setlength{\tabcolsep}{4pt}
\begin{tabular}{llc}
\toprule
 & Parameter & Value \\
\midrule
$K$ & question budget & 5 / 10 / 15 \\
$N$ & candidates per (re)generation & 10 \\
$T_{\mathrm{div}}$ & diverse-sampling temperature & 0.7 \\
$\theta$ & entropy floor & 0.5 \\
\bottomrule
\end{tabular}
\caption{Hyperparameters of our method. $N$ and $T_{\mathrm{div}}$ are swept in
the main paper's sensitivity ablation; the rest are fixed across all
experiments.}
\label{tab:hyperparams}
\end{table}

\paragraph{Baselines.}
All baselines use the same budget $K$, the same proxy, and the same output
protocol (final spec regenerated from the clarified intent). Best-of-N and
Self-Consistency generate 10 candidate questions per round, matching our pool
size. ATD is evaluated at $K\in\{5,10\}$ because of its cost
(\S\ref{app:cost-atd}).


\section{Evaluation Metrics}
\label{app:metrics}

We evaluate a generated specification against the ground-truth reference along
the two axes of IaC ambiguity: whether the infrastructure has the right
\textbf{structure}, and whether its concrete \textbf{values} are exactly right.
Both metrics operate on specs normalized to a common label namespace
(\S\ref{app:norm}). \textbf{Structure} scores the resource/topology graph via a
normalized Graph Edit Distance (\S\ref{app:ged}); \textbf{Res+Attr F1} is a
deliberately strict, exact-match score over concrete resource instances and
attribute values (\S\ref{app:raf1}).

\subsection{Preliminaries}
\label{app:prelim}

\paragraph{Spec representation.}
Each specification $s$ consists of three fields: \textbf{Resources}
$\Rset(s)$, a mapping from a label to a Terraform resource address (e.g.,
\code{main} $\mapsto$ \code{aws\_vpc.main}); \textbf{Topology} $\Tset(s)$, a
mapping from each label to the labels it depends on; and \textbf{Attributes}
$\Aset(s)$, a mapping from each label to a key--value dictionary of
configuration parameters (e.g., \code{cidr\_block}, \code{instance\_type}).
This is the operational form of the triple $s=(R,T,A)$ of
Eq.~\eqref{eq:spec-triple}.

\paragraph{Resource-type extraction.}
Given a Terraform address $a$, the resource type $\rtype(a)$ is its
provider-type prefix: \code{aws\_vpc} for \code{aws\_vpc.main}. For data
sources the \code{data.} prefix is part of the type (e.g.,
\code{data.aws\_iam\_policy\_document}).

\subsection{Label Normalization}
\label{app:norm}

LLMs choose arbitrary instance names (\code{main\_vpc}, \code{vpc1},
\code{v}) for the same resource type, so two specs describing the same design
may differ only in these names. Such syntactic variants should compare as
equal; architecturally distinct configurations should not. We therefore apply
a deterministic label normalization to both the generated and the reference
spec before scoring.

\paragraph{Procedure.}
Given a spec, we (1)~coerce every resource address to a string (some models
emit an address as a singleton list, e.g., \code{["aws\_vpc.main"]});
(2)~group resource labels by their extracted type $\rtype(\cdot)$; (3)~sort
the labels within each type group, for determinism; and (4)~reassign each
label a type-indexed name (e.g., \code{aws\_vpc\_0}, \code{aws\_subnet\_0},
\code{aws\_subnet\_1}), rewriting the address accordingly
(\code{aws\_vpc.aws\_vpc\_0}; data sources keep their \code{data.} prefix).
Topology dependencies and attribute maps are re-keyed through the
resulting label map. The output is a spec whose label set depends only on the
resource types present and their multiplicities.

\paragraph{Effect on each metric.}
Both metrics compare resource types rather than instance names---GED matches
nodes by type, and Res+Attr F1 keys every item by type---so the scores are
already invariant to renaming. Normalization canonicalizes malformed
addresses, fixes one deterministic label scheme for both specs, and keeps the
node alignments reported by GED readable.

\subsection{Structure Score (GED)}
\label{app:ged}

\paragraph{Graph construction.}
Each normalized spec is converted to a labeled directed graph $G=(V,E)$: each
resource label becomes a node $v$ with
$\mathrm{type}(v)=\rtype(\Rset[\ell])$, and each topology dependency
$(\ell \to d)$ becomes a directed edge, added only when both endpoints exist
as nodes.

\paragraph{Cost model.}
We compute the Graph Edit Distance between the reference graph
$G_{\text{ref}}$ and the generated graph $G_{\text{gen}}$ under unit costs:
node and edge insertions and deletions cost~$1$; substituting a node by one of
the same type is free, by one of a different type costs~$1$; edge
substitutions are free, as edges carry no labels. 
This penalizes missing or extra resources and incorrect wiring in a single quantity, unlike type-count schemes that score resources and topology independently.

\paragraph{Computation.}
GED is computed with NetworkX's anytime edit-path search; we keep the best
path found within a 30-second
per-task timeout. The distance is bounded
above by
$D=|V_{\text{ref}}|+|E_{\text{ref}}|+|V_{\text{gen}}|+|E_{\text{gen}}|$, the
cost of deleting every reference element and inserting every generated one.

\paragraph{Normalized score.}
The raw distance is normalized by the total graph size $D$ and inverted:
\begin{equation}
  S_{\text{struct}} \;=\; \max\!\left(0,\; 1-\frac{\mathrm{GED}(G_{\text{ref}},G_{\text{gen}})}{D}\right),
\end{equation}
so $S_{\text{struct}}\in[0,1]$, with $1.0$ an exact structural match; two
empty graphs score $1.0$. The
score reads as the fraction of the maximal structural distance that the
generation closes---a score of $0.55$ means just over half the distance to the
reference is closed. It measures equivalence to the reference, not
deployability; the main paper's deployability ablation measures the latter
via \code{terraform plan} pass rates.

\subsection{Res+Attr F1 (Exact-Match)}
\label{app:raf1}

Structure captures graph shape but not concrete configuration values. Res+Attr
F1 asks the complementary question: how much of the ground-truth spec---its
resource instances and its meaningful attribute values---did the generation
reproduce exactly? It requires no graph matching; it is computed by
occurrence-aware multiset (bag) intersection.

\paragraph{Bag construction.}
Each normalized spec $s$ is reduced to one occurrence-aware bag
$\mathcal{B}(s)$ of items. For every resource label $\ell$ with address $a$:
\begin{itemize}[leftmargin=1.4em,nosep]
  \item add one \textbf{resource item} $(\textsf{R},\,\rtype(a))$;
  \item flatten the attribute map $\Aset(s)[\ell]$ to dotted-key/value leaves
        (below), and for each retained scalar value under key $k$, add one
        \textbf{attribute item} $(\textsf{A},\,\rtype(a),\,k,\,\hat{v})$, where
        $\hat{v}$ is the canonicalized value.
\end{itemize}
Because the bag is a multiset, multiplicity counts: three \code{aws\_subnet}
instances in the reference but two in the generation yield at most two matched
resource items and one missing. Attribute items are keyed by resource type, so
an attribute value can only match if its resource type was produced---attribute
credit is conditional on resource coverage.

\paragraph{Attribute flattening.}
Attribute values are flattened to \code{dotted.key} $\to$ scalar (or list of
scalars) leaves. On the reference side, whose attributes are extracted from
the reference program's \code{terraform plan} output, expression wrappers are
unwrapped
(\code{\{constant\_value: x\}} $\to$ \code{x}) and cross-resource pointers
(\code{\{references: \dots\}}) are dropped: references encode topology, which
Structure already scores. Nested blocks (lists of dicts) flatten under a
shared dotted prefix; a list of scalars is kept as a leaf and later exploded
into one item per element; \code{null} leaves are dropped on both sides.

\paragraph{Value canonicalization.}
Each scalar is canonicalized before matching: booleans map to
\code{bool:<true|false>}, numbers to \code{num:<float>} (so \code{1},
\code{1.0}, and \code{"1"} agree), and any other value to its case-folded,
whitespace-stripped string. There is no embedding or fuzzy credit: values that
are close in form can be incompatible in behavior (\code{10.0.0.0/16} vs.\
\code{10.1.0.0/16} do not match).

\paragraph{Exclusions.}
Two families of items are dropped from both bags because they carry no
cross-spec matching signal: (1)~\emph{free-form and name-like keys}, identified
by a leaf-name denylist---leaf $\in\{$\code{name}, \code{description},
\code{comment}, \code{tags}, \code{arn}, \code{id}$\}$ or a leaf ending in
\code{\_name}, \code{\_arn}, or \code{\_id}---since user-chosen names, human
descriptions, and opaque IDs legitimately differ between a correct generation
and the reference; and (2)~generated-side \emph{reference interpolations},
i.e.\ values containing \code{\$\{\dots\}}, the generated analogue of the
reference's dropped \code{\{references\}} pointers.

\paragraph{Score.}
Let $\mathcal{B}_{\text{ref}}$ and $\mathcal{B}_{\text{gen}}$ be the reference
and generated bags. The matched count is the multiset intersection
\begin{equation}
  m \;=\; \bigl|\mathcal{B}_{\text{ref}}\cap\mathcal{B}_{\text{gen}}\bigr|
       \;=\; \sum_{i} \min\!\bigl(\mathcal{B}_{\text{ref}}(i),\,\mathcal{B}_{\text{gen}}(i)\bigr),
\end{equation}
and the score is the exact-match F1 of the combined resource-and-attribute
bags:
\begin{equation}
  P=\frac{m}{|\mathcal{B}_{\text{gen}}|},\qquad
  R=\frac{m}{|\mathcal{B}_{\text{ref}}|},\qquad
  S_{\text{R+A F1}}=\frac{2PR}{P+R}.
\end{equation}
An empty generated bag scores $0$. Resources and attributes share one bag, so
a single F1 rewards both provisioning the right resource instances and setting
the right values. The score is not comparable to soft or embedding-based
attribute metrics: values must match exactly after canonicalization.

\subsection{Scoring Against a Single Reference}
\label{app:single-ref}

Each task has one verified reference configuration---a valid target, not the
unique one. Both metrics give partial credit accordingly: Structure is
normalized and rewards overlap, so a valid alternative architecture does not
score $0$ but scores in proportion to how much of the reference graph it
reproduces,
and Res+Attr F1 credits every exactly-matching resource instance and attribute
value.
Absolute scores are still bounded by the single-reference design: many valid
programs satisfy the same intent, and Res+Attr F1 is intentionally strict
about exact values (a different but workable CIDR scores $0$ on that item),
which depresses absolute numbers for every method. Since all methods are
evaluated against the same reference under the same procedure, the relative
comparison carries the signal.

\section{Computational Cost}
\label{app:cost}

We report per-task token usage and wall-clock time for every method.
Tokens/task is the mean total number of LLM tokens (prompt and completion)
consumed per task; Time/task is
the mean per-task runtime. The quality scores of these same runs are in the
main paper.

\subsection{Cost Across Backbone Models}
\label{app:cost-models}

Table~\ref{tab:cost-models} gives the per-task cost of our method and the
three lightweight baselines on all five backbones at $K{=}5$, the budget used
in the main paper's cross-model comparison. Our method costs more than the three lightweight baselines---the
price of maintaining and regenerating the candidate pool---and the gap is
similar across backbones. ATD is compared separately in \S\ref{app:cost-atd}.

\begin{table}[t]
\centering
\small
\setlength{\tabcolsep}{3pt}
\begin{tabular}{llrr}
\toprule
Model & Method & Tokens/task & Time/task \\
\midrule
\multirow{4}{*}{Claude Opus 4.6}
 & Ours              & 35.8k & 75s \\
 & Direct Q Gen.     & 2.5k  & 16s \\
 & Best-of-N         & 10.0k & 64s \\
 & Self-Consistency  & 10.6k & 43s \\
\midrule
\multirow{4}{*}{GPT-4.1-mini}
 & Ours              & 35.0k & 58s \\
 & Direct Q Gen.     & 1.9k  & 7s  \\
 & Best-of-N         & 8.4k  & 29s \\
 & Self-Consistency  & 8.8k  & 24s \\
\midrule
\multirow{4}{*}{Qwen3-Coder NEXT}
 & Ours              & 37.8k & 73s \\
 & Direct Q Gen.     & 1.9k  & 7s  \\
 & Best-of-N         & 9.3k  & 27s \\
 & Self-Consistency  & 10.0k & 20s \\
\midrule
\multirow{4}{*}{Llama-4-Scout}
 & Ours              & 33.9k & 47s \\
 & Direct Q Gen.     & 1.8k  & 4s  \\
 & Best-of-N         & 8.2k  & 15s \\
 & Self-Consistency  & 8.4k  & 12s \\
\midrule
\multirow{4}{*}{GPT-4o-mini}
 & Ours              & 31.8k & 28s \\
 & Direct Q Gen.     & 1.8k  & 3s  \\
 & Best-of-N         & 8.2k  & 14s \\
 & Self-Consistency  & 8.6k  & 13s \\
\bottomrule
\end{tabular}
\caption{Per-task cost across five backbone LLMs at $K{=}5$ (Ambig-IaC, 300
tasks).}
\label{tab:cost-models}
\end{table}

\subsection{Cost vs.\ Interaction Budget}
\label{app:cost-rounds}

Table~\ref{tab:cost-rounds} shows how cost scales with the question budget
$K\in\{5,10,15\}$ on GPT-4o-mini. Our token cost grows roughly linearly in
$K$: each round adds one disagreement-driven question and, when the pool is
exhausted, a conditioned regeneration. Direct
question generation is nearly flat, since each round adds only one short
prompt; Self-Consistency grows fastest among the baselines because each round
generates, embeds, and clusters a fresh batch of candidate questions.

\begin{table}[t]
\centering
\small
\begin{tabular}{lrrr}
\toprule
Method & $K{=}5$ & $K{=}10$ & $K{=}15$ \\
\midrule
Ours              & 31.8k / 28s & 61.8k / 50s & 93.8k / 69s \\
Direct Q Gen.     & 1.8k / 3s   & 2.0k / 4s   & 2.2k / 6s   \\
Best-of-N         & 8.2k / 14s  & 18.0k / 31s & 30.0k / 48s \\
Self-Consistency  & 8.6k / 13s  & 25.9k / 38s & 55.4k / 91s \\
\midrule
\emph{Direct} ($K{=}0$) & \multicolumn{3}{c}{0.7k / 1s} \\
\bottomrule
\end{tabular}
\caption{Per-task cost (Tokens/task~/~Time/task) vs.\ interaction budget $K$
(GPT-4o-mini, Ambig-IaC). \emph{Direct} is the no-clarification reference.}
\label{tab:cost-rounds}
\end{table}

\subsection{Comparison to ATD}
\label{app:cost-atd}

ATD scores freely generated candidate questions over a
question~$\times$~solution grid, so its per-round cost scales with the product
of the two; ours scales with the candidate pool alone, since questions are
derived from computed diffs rather than generated and re-scored.
Table~\ref{tab:cost-atd} shows what this means in practice on GPT-4o-mini. At
$K{=}5$, ATD spends $4.1\times$ our tokens and $10.8\times$ our wall-clock per
task---the $4\times$/$11\times$ figures quoted in the main paper---while
trailing our Structure score. At $K{=}10$, ATD's recorded token figure is
inflated by prompt-cache reads; even excluding them it spends $3\times$ our
tokens, and its wall-clock remains an order of magnitude higher.

\begin{table}[t]
\centering
\small
\begin{tabular}{llrr}
\toprule
$K$ & Method & Tokens/task & Time/task \\
\midrule
\multirow{2}{*}{5}
 & Ours & 31.8k & 28s \\
 & ATD  & 129.1k & 303s \\
\midrule
\multirow{2}{*}{10}
 & Ours & 61.8k & 50s \\
 & ATD  & 1451.6k$^{\dagger}$ & 512s \\
\bottomrule
\end{tabular}
\caption{Per-task cost of ATD vs.\ ours (GPT-4o-mini, Ambig-IaC).
$^{\dagger}$Includes prompt-cache reads ($\sim$4.1M cached tokens per task);
$188.0$k excluding them.}
\label{tab:cost-atd}
\end{table}

\section{Output-Protocol Ablation}
\label{app:output}

Our method differs from the baselines in two places: which questions it asks
and how it emits the final specification. The baselines share one output
protocol---after $K$ rounds, the accumulated Q\&A is folded into a
natural-language \emph{clarified intent}, and the spec generator produces the
final spec from it. Ours instead returns the best surviving pool candidate, falling back to a single conditioned generation only when nothing survives. This ablation separates the two.

\paragraph{Setup.}
\textbf{Ours-BaselineOut} keeps our question selection unchanged and replaces
only the output stage: after the interaction, the same generator receives the
original task and the accumulated Q\&A history, constructs a clarified intent,
and generates the final specification from it---exactly the baselines'
protocol. The comparison is at $K{=}15$ on GPT-4o-mini; the largest budget
gives the regeneration protocol the longest history to work from.

\paragraph{Results.}
Table~\ref{tab:output} reports the ablation. Ours-BaselineOut still beats every
baseline on both metrics, so disagreement-driven question selection helps on
its own, independent of the output stage. The full method adds a further gain,
mostly in Structure: a surviving pool candidate is
consistent with the answers by construction---generated under them and pruned
by them---while a spec regenerated from free text has no such guarantee. Read
as a pair, the two gaps separate the contributions:
Ours-BaselineOut vs.\ the baselines isolates question selection; Ours vs.\
Ours-BaselineOut isolates the output stage.

\begin{table}[t]
\centering
\small
\begin{tabular}{lrr}
\toprule
Method & Struct.\ (\%) & R+A F1 (\%) \\
\midrule
\textbf{Ours}               & \textbf{54.84} & \textbf{26.53} \\
Ours-BaselineOut            & 49.07~(\dnar{10.5\%}) & 26.01~(\dnar{2.0\%}) \\
\midrule
Direct Q Generation         & 46.31~(\dnar{15.5\%}) & 18.45~(\dnar{30.5\%}) \\
Best-of-N                   & 43.12~(\dnar{21.4\%}) & 19.52~(\dnar{26.4\%}) \\
Self-Consistency            & 38.19~(\dnar{30.4\%}) & 16.12~(\dnar{39.2\%}) \\
\bottomrule
\end{tabular}
\caption{Output-protocol ablation (GPT-4o-mini, $K{=}15$, Ambig-IaC).
\textbf{Ours-BaselineOut} keeps our question selection but emits the final spec
via the baseline clarified-intent regeneration. Parentheses give the relative
drop from \textbf{Ours}.}
\label{tab:output}
\end{table}

\section*{Prompts}

This section reproduces the key prompts used in our pipeline, shown as
full-width boxes on this and the following pages; text in angle brackets,
\code{<like this>}, marks values substituted at run time.

\subsection*{Specification Generation}

The system prompt and the accompanying user message are shown in the
\emph{Specification generation} boxes; \code{Feedback / issue} is omitted
from the user message before the first answer.

\subsection*{Clarification-Question Generation}

\textsc{Verbalize} (\S\ref{app:qgen}) sends the \emph{Question generation}
system prompt with \code{\{n\}} instantiated to $1$; the \code{keep(...)}
labels become the keep sets $\mathcal{P}^{\textsf{yes}}$ and
$\mathcal{P}^{\textsf{no}}$. The diff summary supplied in the user message
renders the shortlist $F^\ast$ per axis (empty sections and ID-like
attribute keys omitted).

\subsection*{Simulated User (Proxy)}

The proxy (GPT-4o-mini, temperature $0$ for every method and backbone)
answers strictly from the ground-truth configuration, appended to the
system prompt as \code{Reference configuration: <ground-truth main.tf>};
each question arrives as its own user message.

\subsection*{Interaction-History Conditioning}

The accumulated Q\&A is folded into a single \emph{clarified intent}
string used to condition (re)generation (\S\ref{app:gen}).

\subsection*{Spec-to-Terraform Conversion}

The deployability ablation converts each method's final spec into a
\code{main.tf} and runs \code{init}$\,\to\,$\code{validate}$\,\to\,$%
\code{plan}. Emission user message:
\code{Clarified intent: <clarified intent>} then
\code{Infrastructure spec: <spec JSON>}; failing tasks get one repair pass
whose user message adds the failing HCL and the error tail.

\begin{promptbox}{Specification generation (system prompt)}
You are an infrastructure spec generator for AWS Terraform. You are generating ONE POSSIBLE interpretation of an ambiguous infrastructure request. The request can be interpreted in multiple valid ways — your job is to pick a SPECIFIC interpretation that may differ from others.

Focus on STRUCTURAL diversity — vary your choices in:
- **Resources**: Consider different AWS services that could fulfill the same need (e.g., ALB vs NLB, RDS vs Aurora vs DynamoDB, CloudWatch vs S3 for logging, Lambda vs ECS vs EC2)
- **Topology**: Consider different dependency structures and how resources connect (e.g., direct vs through a gateway, single-tier vs multi-tier, which resources reference which)
- **Resource count**: Consider different numbers of resources (e.g., 2 replicas vs 3, single AZ vs multi-AZ)

Output a single JSON object with exactly these keys:

## 'resources':
Maps short label strings to full Terraform addresses in the form 'aws_resource_type.instance_name'. Use real AWS Terraform provider resource types.

## 'topology':
Maps src_label to list of dep_labels (dependency edges). Labels must match keys in 'resources'.

## 'attributes':
Maps each resource label to a dict of key Terraform attributes for that resource.

# Guidelines:
- Focus on structural choices (which resources exist, how they connect) rather than specific IDs or values.
- Do NOT invent specific IDs, ARNs, subnet IDs, AMI IDs, zone IDs, or account numbers.
- For attributes that reference other resources, use Terraform resource references (e.g., "${aws_db_instance.main.id}", "${aws_route53_zone.primary.zone_id}") instead of hardcoded IDs.
- For attributes, include structurally meaningful ones (e.g., engine type, instance_class, record type, source_db_instance_identifier) using resource references where appropriate.
- Include topology edges whenever resources logically depend on each other.
- Follow the user request and any clarification Q&A, but fill in ambiguous structural details with specific choices.

Output ONLY the JSON object, no markdown, no explanations.
\end{promptbox}
\begin{promptbox}{Specification generation (user message)}
Task description: <request>

Feedback / issue: <clarified intent>

This is interpretation <i> of <N>. Make specific, concrete choices that may differ from other interpretations.
\end{promptbox}

\begin{promptbox}{Question generation (system prompt)}
You are an AWS infrastructure expert generating clarification questions.

You are given:
1. An ambiguous infrastructure request from a user
2. A summary of structural differences found across multiple candidate interpretations (specs)
3. Each candidate spec is identified by an index (0, 1, 2, ...)

Your job: generate {n} human-readable yes/no clarification questions that resolve the most important structural ambiguities. Each question must include filter labels showing which spec indices would be KEPT for each answer.

Rules:
- Questions must be natural, human-readable language — no JSON, no terraform syntax, no resource IDs.
- Focus on STRUCTURAL questions: which services to use, how they connect, how many of each.
- Do NOT ask about specific attribute values like IDs, ARNs, CIDR blocks, endpoint URLs, names, or regions.
- Each question should discriminate between candidate specs — if all specs agree, don't ask about it.
- IMPORTANT: Each answer option (Yes/No) should keep at least 2 specs when possible. Avoid questions where one answer keeps only 1 spec — those are too aggressive and waste the question budget.
- IMPORTANT: Questions must be precise and non-overlapping. Do not ask two questions about the same topic with slightly different wording.
- Label each question with its dimension: [resource], [topology], or [attribute].

Output format — output ONLY the questions, no code fences, no extra text:

[dimension] Question text here?
Yes -> keep(0,2,4)
No -> keep(1,3)

[dimension] Another question?
Yes -> keep(0,1)
No -> keep(2,3,4)
\end{promptbox}

\begin{promptbox}{Question generation (user message)}
Infrastructure request: <request>

<diff summary>

Generate exactly 1 clarification question.
\end{promptbox}
\begin{promptbox}{Diff-summary format (input to the question LLM)}
=== <N> Candidate Specs ===
Spec 0:
  Resources: <comma-separated resource types>
  Connections: <src_type -> dep_type, ...>
Spec 1:
  ...

=== Resource Disagreements (specs disagree on whether these exist) ===
- <resource_type>: YES in specs [<ids>], NO in specs [<ids>]

=== Topology Disagreements (specs disagree on these connections) ===
- <src_type> -> <dep_type>: YES in specs [<ids>], NO in specs [<ids>]

=== Attribute Disagreements (structural, non-ID values only) ===
- <resource_type>.<key>: spec <id>: <value>, spec <id>: <value>, ...
\end{promptbox}

\begin{promptbox}{Simulated user (system prompt)}
You are an expert in AWS cloud infrastructure and cloud configuration.
You will be given a cloud configuration and true/false questions about the configuration.
You will need to answer the questions strictly based on the configuration.

Guidelines:
- Only answer with yes or no.
- You must strictly follow the reference configuration to answer the question.

Example 1:
Task: Create an AWS resource to help me store images that I want to display on my website

Reference configuration:
provider "aws" {
  region = "us-east-1"
}

resource "aws_s3_bucket" "example-bucket" {
  bucket = "test-bucket"
}

Question 1: Do you need audit logging for access to the stored images?
Answer: No
Reason: The reference configuration does not contain any audit logging for access to the stored images.

Question 2: Do you want an S3 bucket for storing the images?
Answer: Yes
Reason: The reference configuration contains an S3 bucket for storing the images.

Question 3: Do you require any specific backup or recovery options for the stored images?
Answer: No
Reason: The reference configuration does not contain any specific backup or recovery options for the stored images.

End of example.
\end{promptbox}

\begin{promptbox}{Clarified-intent template}
Task: <original request>

Q: <question 1>
A: <Yes | No>

Q: <question 2>
A: <Yes | No>
...
\end{promptbox}

\begin{promptbox}{Spec-to-Terraform emission (system prompt)}
You are an AWS Terraform expert. Given an infrastructure spec (resources, topology, attributes) and the user's clarified intent, produce a single main.tf that `terraform plan` will accept.

Requirements:
- Output ONLY valid Terraform HCL. No markdown fences, no commentary.
- Include a terraform { required_providers { aws = ... } } block and a provider "aws" block (region = "us-east-1" unless intent says otherwise).
- Use the spec's labels as resource names where reasonable.
- Fill required attributes with sensible defaults; favor a plan-clean configuration over completeness.
- Do not reference undeclared resources or variables.
\end{promptbox}

\begin{promptbox}{Error-guided repair pass (system prompt)}
You are an AWS Terraform expert. The Terraform HCL below failed `terraform validate` or `terraform plan`. Fix it so both pass.

Requirements:
- Output ONLY valid Terraform HCL. No markdown fences, no commentary.
- Preserve the user's clarified intent and the resource set from the infrastructure spec where possible.
- Address the specific terraform error shown. Add missing required attributes, fix invalid references, and drop unsupported/hallucinated ones.
- Keep the terraform { required_providers { aws = ... } } block and a provider "aws" block.
\end{promptbox}